\documentclass[aps,prb,superscriptaddress,preprint]{revtex4-1} 

\usepackage[colorlinks=true,linkcolor=blue,citecolor=blue,urlcolor=blue]{hyperref}

\usepackage{setspace} 
\usepackage{graphicx}
\usepackage{amsmath}
\usepackage{color}
\usepackage{amsmath}
\usepackage{amssymb}
\usepackage{verbatim}
\usepackage{latexsym}
\usepackage{enumerate} 
\usepackage{bm} 
\usepackage{ulem} 
\usepackage[caption=false]{subfig}
\usepackage{multirow}
\usepackage{booktabs}
\usepackage{lineno}

\begin{document}

\title{%
Monolayer fullerene networks as photocatalysts for overall water splitting
}

\author{Bo Peng}
\email{bp432@cam.ac.uk}
\affiliation{Theory of Condensed Matter Group, Cavendish Laboratory, University of Cambridge, J.\,J.\,Thomson Avenue, Cambridge CB3 0HE, United Kingdom}

\date{\today}

\begin{abstract}
Photocatalytic water splitting can produce hydrogen in an environmentally friendly way and provide alternative energy sources to reduce global carbon emissions. Recently, monolayer fullerene networks have been successfully synthesized\ [Hou \textit{et al., Nature} \textbf{2022}, 606, 507], offering new material candidates for photocatalysis because of their large surface area with abundant active sites, feasibility to be combined with other 2D materials to form heterojunctions, and the C$_{60}$ cages for potential hydrogen storage. However, efficient photocatalysts need a combination of a suitable band gap and appropriate positions of the band edges with sufficient driving force for water splitting. In this study, I employ semilocal density functional theory and hybrid functional calculations to investigate the electronic structures of monolayer fullerene networks. I find that only the weakly screened hybrid functional, in combine with time-dependent Hartree-Fock calculations to include the exciton binding energy, can reproduce the experimentally obtained optical band gap of monolayer C$_{60}$. All the phases of monolayer fullerene networks have suitable band gaps with high carrier mobility and appropriate band edges to thermodynamically drive overall water splitting. In addition, the optical properties of monolayer C$_{60}$ are studied, and different phases of fullerene networks exhibit distinct absorption and recombination behavior, providing unique advantages either as an electron acceptor or as an electron donor in photocatalysis.
\end{abstract}

\flushbottom

\maketitle

\section*{Introduction}

The energy consumption of fossil fuels is the main source of global carbon emissions\,\cite{Bhatia2022}. As an alternative, hydrogen can be burnt in the presence of oxygen and produce only water, supporting mitigation of CO$_2$ emissions. Photocatalysis can decompose water into hydrogen and oxygen using light, providing a low-cost approach for the green production of hydrogen. Photocatalytic water splitting has been extensively studied since the discovery of electrochemical photolysis of water in TiO$_2$ in 1972\,\cite{Fujishima1972,Deak2011,Scanlon2013,Pfeifer2013,Ju2014,Mi2015,Zhang2015n,Deak2016,Chiodo2010,Li2020a}. However, due to the wide band gap of $3.0-3.2$ eV in TiO$_2$, only the ultraviolet part of the solar spectrum can be harnessed. To maximize the photocatalytic efficiency, a water-splitting material need to ($i$) absorb the light effectively to generate enough electron-hole pairs; ($ii$) separate the generated carriers on the surface; and ($iii$) overcome the potential barrier of the reaction. For ($i$) and ($iii$), a compromise of the band gap is needed to harness the photon energy effectively while fulfilling the requirements of the band edges to facilitate the redox reaction of water. As a result, an optimal band gap around 2 eV is required, and the band edges must span the redox potential\,\cite{LeBahers2014,Wang2019,Brlec2022}. For ($ii$), a type-II band alignment can spontaneously separate the electrons and holes. Based on these requirements, a variety of candidate materials have been proposed for efficient water splitting\,\cite{Norskov2004,Rossmeisl2007,Zhang2007,Zhang2011,Suzuki2012,Jiang2013,Xu2013b,Zhu2013,Zheng2015,Qiao2018,Yang2019,Ju2020,Nakada2021,Luo2022,Wang2022,Fu2022}. Among all the candidates, carbon nanomaterials exhibit high physical stability and rich redox chemistry\,\cite{Pan2020,Yao2020}. In particular, fullerene, the cage structure of C$_{60}$\,\cite{Kroto1985}, displays high quantum efficiency in photocatalytic reactions because of their large surface area, abundant micropores, increased surface active sites and efficient electron transport properties\,\cite{Barker2016,Chen2018,Ma2019,Arivazhagan2019}. In photocatalysis, C$_{60}$ can enhance the photocatalytic activity via different mechanisms: it can work as an electron acceptor owing to rapid carrier separation\,\cite{Kamat1994,Yu2011a,Youssef2018,Ma2019}, or as an energy transfer mediator\,\cite{Panagiotou2010}, or as an electron donor due to high photosensitivity\,\cite{Song2016b}. In addition, for composite materials, the introduction of fullerene results in better crystallization by reducing the defects\,\cite{Arivazhagan2019}, and can also improve the stability of the composites\,\cite{Fu2008a,Du2016}, which further enhance the photocatalytic efficiency. Most interestingly, C$_{60}$ itself is a promising hydrogen storage material\,\cite{Zhao2005,Yoon2008,Wang2009,Wang2012d,Durbin2016}, and photocatalytic water splitting using fullerene provides a convenient approach to produce and store hydrogen at the same time.

Recently, a 2D material composed of covalently bonded fullerene network structures has been synthesized, with two configurations obtained: a few-layer quasi-tetragonal phase (qTP) and a monolayer quasi-hexagonal phase (qHP)\,\cite{Hou2022}. The various structural phases of monolayer fullerene networks can be combined with other 2D materials to form type-II van der Waals heterostructures\,\cite{Ozcelik2016,Hu2017,Yang2020a}, which can efficiently separate carriers between individual layers. In addition, the band alignment in 2D heterostructures can be further controlled by external strain because of the mechanical flexibility of 2D materials\,\cite{Tromer2022,Yu2022,Peng2017a}. Compared to heterostructures using C$_{60}$ molecules where the low C$_{60}$ content is not periodically bounded at the edge of the other 2D material\,\cite{Guan2018}, heterostructures using monolayer polymetric fullerene has a smooth microscopic surface with uniform periodic C$_{60}$ networks, which provides higher crystallinity with higher C$_{60}$ concentrations and consequently increases the photocatalytic activity. Compared to other 2D materials\,\cite{Kim2015d,Zhang2016e,Yu2016,Pierucci2016,Hill2016,Zhang2018d,Zhang2018e,Zhang2019d,Fu2020a,Zhang2020a,Han2020,Chen2022}, monolayer C$_{60}$ has larger surface area with more active sites due to the quasi-0D network structures of C$_{60}$ cages. Additionally, monolayer C$_{60}$ exhibits good thermodynamic stability and high carrier mobility\,\cite{Hou2022}. All these physical/chemical properties render monolayer fullerene networks a promising candidate for photocatalytic water splitting. However, all theoretical calculations underestimate the band gap of monolayer C$_{60}$ by at least 10\%\,\cite{Tromer2022,Yu2022,Yuan2022}, and a correct description of the band structures is the prerequisite for exploring the band edge positions for water splitting or the optical absorption for photocatalysis.



In this paper, the electronic structures of monolayer qTP and qHP fullerene networks are investigated using semilocal density functional theory (DFT) and hybrid functional calculations. By examining the band gap and exciton binding energy, I find that the electronic structures and optical properties of monolayer C$_{60}$ can only be describe correctly by weakly screened hybrid functional. The band gaps of monolayer fullerene are around $1.67-1.88$ eV, and the band edge positions provide sufficient driving forces for overall water splitting. In addition, monolayer fullerene networks possess high carrier mobility that can effectively transfer the photoexcited electrons and holes. Furthermore, the carrier recombination in qTP C$_{60}$ is suppressed by weak optical transition, leading to efficient carrier separation as an electron acceptor. On the other hand, the strong optical absorption in qHP C$_{60}$ can provide a large amount of electrons for hydrogen evolution, making it promising as an electron donor. These results indicate that monolayer fullerene networks are promising as efficient overall water splitting photocatalysts.

\section*{Methods}

All crystal structures of monolayer fullerene networks are optimized using the PBEsol functional\,\cite{Perdew2008} as implemented in {\sc vasp}\,\cite{Kresse1996,Kresse1996a}. A plane-wave cutoff of 800 eV is used with a {\bf k}-mesh of $5\times5$ and $3\times5$ for qTP and qHP C$_{60}$ respectively. During the structural relaxation, an energy convergence criterion of 10$^{-6}$ eV and a force convergence criterion of 10$^{-2}$ eV/\AA\ are enforced. To mimic the 2D monolayers with 3D periodic boundary conditions, an interlayer vacuum spacing larger than 17 \AA\ is used to eliminate interactions between adjacent unit cells along the $c$ direction.

The electronic structures of qTP and qHP C$_{60}$ are calculated using the screened hybrid functional HSE\,\cite{HSE1,HSE2,HSE3,Schimka2011}. Using the HSE wavefunctions, the partial (band decomposed) charge density is calculated for the top valence and bottom conduction bands at selected {\bf k}-points. The transport properties are calculated based on the HSE eigenenergies and eigenstates in a {\bf k}-mesh of $8\times8$ ($5\times8$) for qTP (qHP) C$_{60}$, which is further interpolated using an interpolation factor of 100. The scattering rates for acoustic deformation potential and ionized impurity scattering are calculated using the {\sc amset} package\,\cite{Ganose2021}. The deformation potential are calculated for anisotropically contracted (-0.5\%) and expanded (+0.5\%) lattice, and the elastic tensor coefficients (including ionic relaxations) are computed using the finite differences method\,\cite{LePage2002,Wu2005}. For ionized impurity scattering, the static dielectric constant is calculated from density functional perturbation theory\,\cite{Gajdos2006}.

When computing the optical properties, the thickness-independent absorbance $A (\omega)$ is calculated from the imaginary part of the dielectric function $\epsilon_2 (\omega)$\,\cite{Bechstedt2012,Matthes2013,Matthes2013a}
\begin{equation}
    A (\omega) = \frac{\omega}{c}  L \epsilon_2 (\omega),
\end{equation}
where $\omega$ is the photon frequency, $c$ is the speed of light and $L$ is the distance between the 2D sheets. The absorbance in the independent particle picture\,\cite{Gajdos2006} is calculated using the hybrid-functional electronic structures. To include the excitonic effects, time-dependent Hartree-Fock (TDHF) calculations are performed on top of the HSE eigenenergies and eigenstates using the Casida equation that includes couplings among the group of resonant/antiresonant two-orbital states\,\cite{Sander2017}. The exciton eigenenergies and their corresponding oscillator strengths can be obtained directly from the Casida equation\,\cite{Sander2017}. The exciton binding energy is then computed as the difference between the eigenenergy in the independent particle picture and the exciton eigenenergy. The Tamm-Dancoff approximation is used as the exciton eigenenergies calculated within and beyond this approximation\,\cite{Sander2015} only have a difference smaller than 5 meV. In 2D materials, the exciton absorption spectrum calculated from TDHF agrees qualitatively well with the results obtained from the Bethe-Salpeter equation (BSE) on top of the $GW$ calculations\,\cite{Peng2018c}, and TDHF is computationally much less expensive than $GW$ + BSE, especially for large systems such as monolayer fullerene networks. A {\bf k}-mesh of $8\times8$ ($5\times8$) is used for qTP (qHP) C$_{60}$, with the highest eight (sixteen) valence bands and the lowest eight (sixteen) conduction bands included as basis, converging the exciton eigenenergy within 1 meV.

To compute the thermodynamics of water adsorption and redox reactions, a supercell of $2\times2$ and $1\times2$ is used for qTP and qHP C$_{60}$ respectively, with an electronic {\bf k}-point grid of $3 \times 3$. Both the lattice constants and internal atomic coordination are fully relaxed for all the atoms. For hydrogen reduction reaction, geometry optimization always results in top-site-adsorbed hydrogen atom. The lowest energy intermediates are evaluated by comparing hydrogen adsorption on all the symmetry irreducible carbon atoms. The thermal corrections at room temperature, including zero-point energy, entropy and internal thermal energy, are calculated using {\sc vaspkit}\,\cite{Wang2021b}. The vibrational frequencies are computed for both the adsorbed hydrogen atoms and the neighboring carbon atoms within a radius of 2.5 \AA.

\section*{Results and Discussion}

\subsection*{Crystal structures}

\begin{figure*}
\centering
\includegraphics[width=0.75\textwidth]{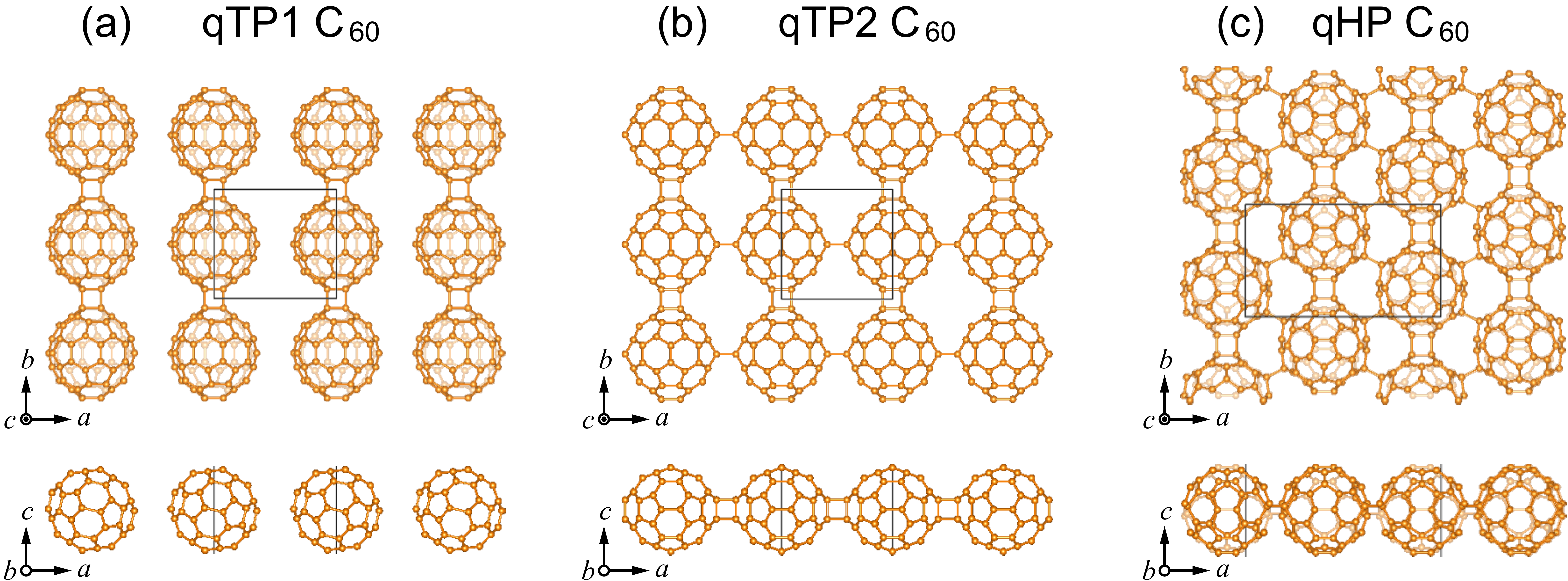}
\caption{
Crystal structures of monolayer (a) qTP1, (b) qTP2 and (c) qHP C$_{60}$ from top and front views.
}
\label{crystal} 
\end{figure*}

The crystal structures of fully relaxed fullerene networks are present in Fig.~\,\ref{crystal}. After geometry optimization, two quasi-tetragonal phases are obtained. One phase, denoted as qTP1, is obtained by structural relaxation starting from the quasi-tetragonal phase consisting of only carbon atoms. The other quasi-tetragonal phase, denoted as qTP2, is obtained by a two-step geometry optimization, which starts with the experimentally reported qTP Mg$_2$C$_{60}$ and then removes the Mg ions before the second relaxation. The two-step structural relaxation is to mimic the experimental procedure to remove the charged ions introduced during synthesis by treatment with hydrogen peroxide to obtain clean single crystals of the carbon polymers\,\cite{Hou2022,Gottfried2022}.

Monolayer qTP1 C$_{60}$ crystallizes in space group $P2/m$ (No.\,10) with lattice parameters $a=10.175$ \AA\ and $b=9.059$ \AA, in which each C$_{60}$ is linked by two neighboring C$_{60}$ cages through two [$2+2$] cycloaddition bonds along the $b$ direction, forming 1D chains of C$_{60}$ cluster cages in Fig.~\,\ref{crystal}(a). The shortest interchain distance between the nearest carbon atoms is 3.065 \AA\ along the $a$ direction, which is much longer than the C$-$C single bonds. The interchain distance is shortened merely by 0.172 \AA\ when including the van der Waals interactions\,\cite{Grimme2010}, therefore the van der Waal forces are neglected in qTP1 C$_{60}$ (for the role of van der Waals forces in the lattice constants of all three phases, see the Supporting Information).
The space group of qTP2 C$_{60}$ is $Pmmm$ (No.\,47), with lattice parameters $a=9.097$ \AA\ and $b=9.001$ \AA. Similar to qTP1 C$_{60}$, the in-plane [$2+2$] cycloaddition bonds connect neighboring C$_{60}$ cages along the $b$ direction in qTP2 C$_{60}$. The difference between qTP1 and qTP2 C$_{60}$ is along the $a$ direction: no bond is formed between neighboring C$_{60}$ chains in qTP1 fullerene along the $a$ direction, whereas each C$_{60}$ cage of qTP2 fullerene connects two neighboring cages along that direction through two out-of-plane [$2+2$] cycloaddition bonds, as demonstrated in Fig.~\,\ref{crystal}(b).
Monolayer qHP C$_{60}$ has a space group of $Pc$ (No.\,7) with lattice parameters $a=15.848$ \AA\ and $b=9.131$ \AA, where each C$_{60}$ is connected to six neighboring C$_{60}$ cages with four C$-$C single bonds along the diagonal lines of the rectangular unit cell and two [$2+2$] cycloaddition bonds along the $b$ direction, as demonstrated in Fig.~\,\ref{crystal}(c). The calculated lattice constants agree well with previous calculations\,\cite{Tromer2022}. The dynamic stability of all three phases is evaluated in the Supporting Information. In addition, thermal stability of monolayer qTP and qHP C$_{60}$ has been confirmed using molecular dynamics simulations, showing that both qTP and qHP C$_{60}$ monolayers can remain stable at temperatures near 800 K\,\cite{Ribeiro2022}, which is in line with the experimental result that monolayer qHP C$_{60}$ does not decompose at 600 K\,\cite{Hou2022}.

\subsection*{Appropriate screening parameter}

To gain insight into the appropriate level of theory to correctly describe the electronic structures and optical properties of the C$_{60}$ monolayers, the electronic and optical band gaps of monolayer qHP C$_{60}$, as well as the exciton binding energy, are calculated using the hybrid functional with different screening parameters $\mu$\,\cite{Perdew1996a,Adamo1999,Ernzerhof1999,Schimka2011}, and then compared with the experimentally determined value. 
In 2D materials, the excitonic effects are stronger than their bulk counterparts due to weaker dielectric screening\,\cite{Latini2017,Peng2018c,Zhang2020a} (for dielectric screening in bulk and monolayer polymeric C$_{60}$, see the Supporting Information). To include exciton binding energy, time-dependent Hartree-Fock calculations are performed on top of different hybrid functionals, which provides qualitatively consistent exciton absorption spectrum compared to $GW$ + BSE and is computationally much less expensive\,\cite{Peng2018c}.

\begin{figure}
\centering
\includegraphics[width=0.4\textwidth]{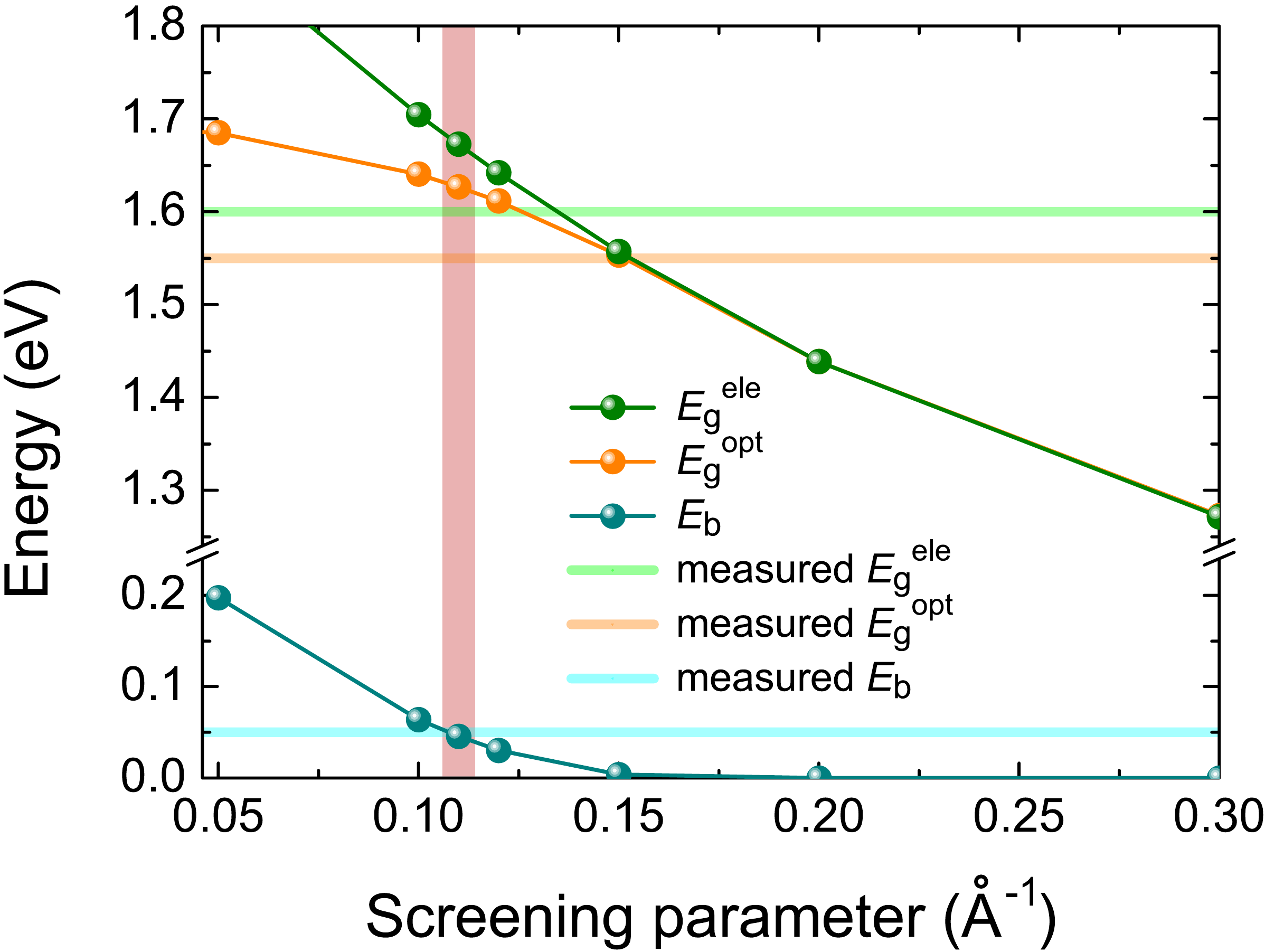}
\caption{
Electronic and optical band gaps of monolayer qHP C$_{60}$ calculated from different screening parameters, as well as the corresponding exciton binding energy.
}
\label{screen} 
\end{figure}

Figure\,\ref{screen} summarizes the electronic band gap $E_{\rm g} ^ {\rm ele}$, optical band gap $E_{\rm g} ^ {\rm opt}$ and exciton binding energy $E_{\rm b}$ of qHP C$_{60}$ computed from different screening parameters $\mu$ (for similar results on qTP C$_{60}$, see the Supporting Information). A screening parameter larger than 0.15 \AA$^{-1}$ not only severely underestimates the electronic band gap $E_{\rm g} ^ {\rm ele}$, but also predicts zero exciton binding energy. For example, the HSEsol (the PBEsol counterpart of the widely used HSE06 with $\mu = 0.2$ \AA$^{-1}$) hybrid functional predicts an electronic band gap of 1.44 eV, and the HSEsol band gap is 10\% narrower compared to the measured gap of 1.6 eV, which can be attributed to an increase in the dielectric screening of HSEsol\,\cite{Savory2016a}. Therefore the HSEsol hybrid functional is inadequate to describe the electronic and optical properties of monolayer fullerene networks, as it tends to overestimate the screening effects in low-dimensional systems and consequently underestimate their band gap and exciton binding energy\,\cite{Peng2018c,Li2021b,Su2022}. 
This is unsurprising because in quasi-0D C$_{60}$ monolayers the screening effects are much weaker than most 2D materials.

Among all the screening parameters below 0.15 \AA$^{-1}$, a screening parameter of 0.11 \AA$^{-1}$ yields an exciton binding energy $E_{\rm b}$ of 0.05 eV, which is in good agreement with the experimental value\,\cite{Hou2022}. The screening length is in excellent agreement with the inverse of the distance between two nearest neighboring buckyballs ($\sim$ 9.1 \AA).  
In addition, it predicts an electronic band gap of 1.67 eV compared to the measured $E_{\rm g} ^ {\rm ele}$ of 1.6 eV, while obtaining a reasonable $E_{\rm g} ^ {\rm opt}$ of 1.62 eV compared to the experimentally obtained 1.55 eV. The tiny discrepancy ($<4.5\%$) may come from temperature effects such as electron-phonon coupling\,\cite{Saidi2016,Monserrat2018a,Wu2018a,Peng2019,Miglio2020}, which are not included in the calculations. Further decreasing the dielectric screening results in an overestimation of both the band gaps and the binding energy. Thus a correct description of the band structures and optical properties can only be obtained by using the weakly screened hybrid functional with $\mu = 0.11$ \AA$^{-1}$ and TDHF on top of the hybrid functional respectively.

\subsection*{Electronic structures and carrier mobilities}

\begin{figure*}[h]
\centering
\includegraphics[width=\textwidth]{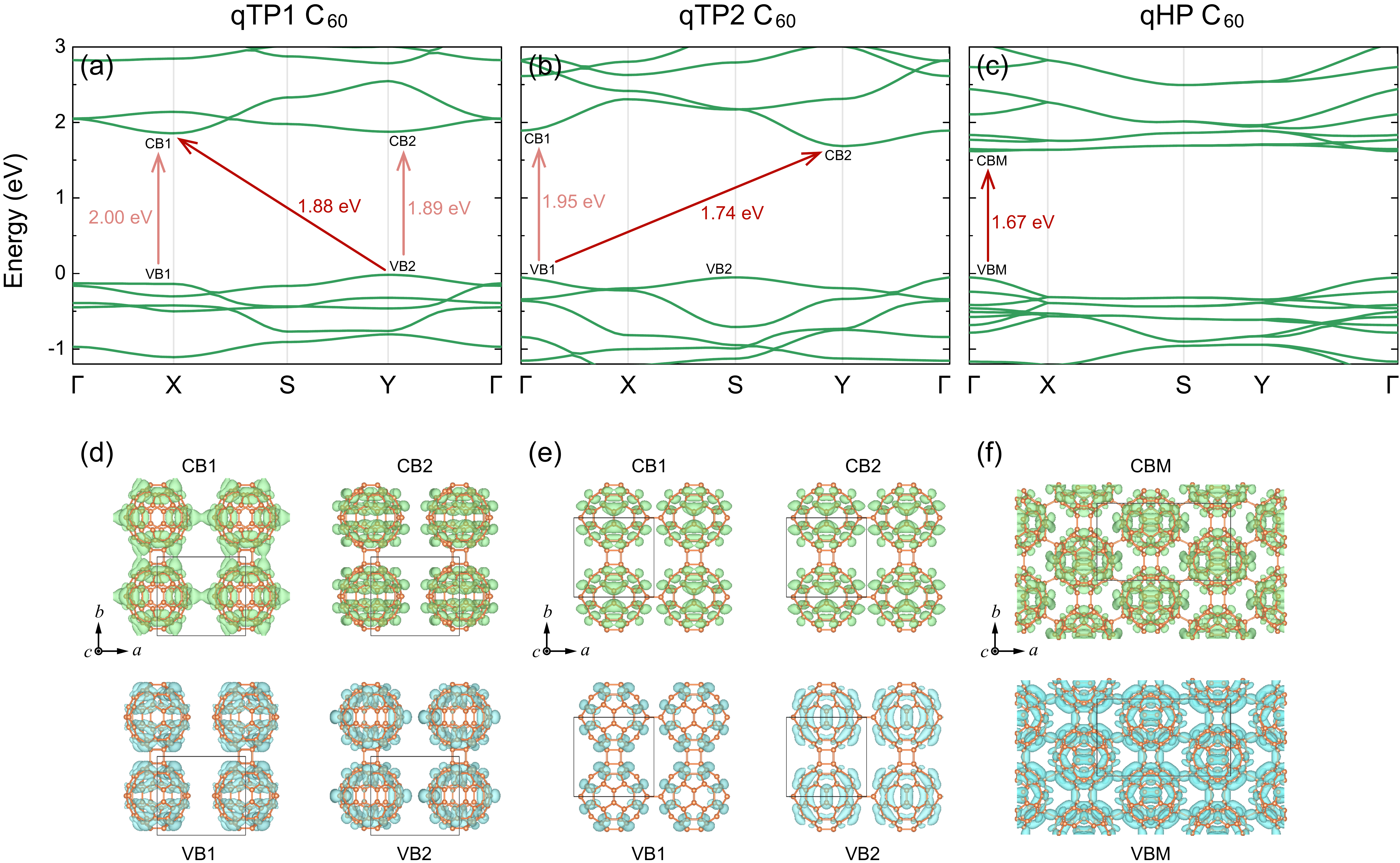}
\caption{
Electronic structures of (a) qTP1, (b) qTP2, and (c) qHP C$_{60}$ calculated with weakly screened hybrid functional ($\mu = 0.11$ \AA$^{-1}$), as well as their corresponding partial charge density of the top valence states and the lowest conduction states in (d)-(f). The default isosurface level is used (0.009 and 0.005 \AA$^{-3}$ for qTP and qHP C$_{60}$ respectively), as implemented in {\sc vesta}\,\cite{vesta}.
}
\label{electron} 
\end{figure*}

Using the weakly screened hybrid functional with $\mu = 0.11$ \AA$^{-1}$, the electronic structures are predicted (for band structures calculated from PBEsol and HSEsol, see the Supporting Information). All three phases have a 2D rectangular Brillouin zone (for details, see the Supporting Information), with high-symmetry points $\Gamma$ (0, 0), X (1/2, 0), S (1/2, 1/2) and Y (0, 1/2). Figure\,\ref{electron}(a) shows the band structures of qTP1 C$_{60}$. The obtained band gap of 1.88 eV is indirect, with the valence band maximum (VBM) at the Y high-symmetry point and the conduction band minimum (CBM) at X. The direct transition energies at X and Y are 2.00 and 1.89 eV respectively. 


To visualize the band edges, the partial charge density for the top valence and bottom conduction bands at X and Y is shown in Fig.\,\ref{electron}(d). The lowest conduction band at X (CB1) is more dispersive and its charge density is more diffuse along both the $a$ and $b$ directions. The highest valence band is flat along $\Gamma-$X, and as expected, the corresponding charge density of the top valence band at X (VB1) is isolated within separated C$_{60}$ cages. Similarly, the top valence states and lowest conduction states at Y, denoted as VB2 and CB2 respectively, are centered around each single C$_{60}$ cage, and such molecular-like character is consistent with their flat bands.

For qTP2 C$_{60}$, the weakly screened hybrid functional predicts an indirect band gap of 1.74 eV with the VBM at $\Gamma$ and the CBM at Y, while the direct transition energy at $\Gamma$ is 1.95 eV. As shown in Fig.\,\ref{electron}(b), the band structures of qTP2 C$_{60}$ show distinct difference from qTP1 C$_{60}$, despite that their lattice parameters are similar. In addition, the charge density of qTP2 C$_{60}$ changes significantly compared to that of qTP1 C$_{60}$. Because the space group of qTP2 C$_{60}$ ($Pmmm$) has more symmetry operations than that of qTP1 C$_{60}$ ($P2/m$), their partial charge density in Fig.\,\ref{electron}(e) is more symmetric than that of qTP1 C$_{60}$. Interestingly, although the lowest conduction band between $\Gamma$ and Y has an energy difference of 0.21 eV, their corresponding partial charge density (denoted as CB1 and CB2 respectively) exhibits no significant difference. In contrast, for the highest valence band, although the energy difference between $\Gamma$ and S is lower than 0.7 meV, their partial charge density (denoted as VB1 and VB2 respectively) is distinct from each other.

Figure\,\ref{electron}(c) depicts the band structures of monolayer qHP C$_{60}$. Monolayer qHP C$_{60}$ possesses a direct band gap at $\Gamma$. The CBM of monolayer qHP C$_{60}$ exhibits flat-band features, and its charge density is molecular-like, as shown in Fig.\,\ref{electron}(f). On the other hand, the charge of the more dispersive VBM is distributed in the entire Brillouin zone, connecting neighboring C$_{60}$ cages via both the C$-$C single bonds and the [$2+2$] cycloaddition bonds. Therefore, holes are expected to diffuse more effectively in qHP C$_{60}$.

\begin{figure*}[h]
\centering
\includegraphics[width=\textwidth]{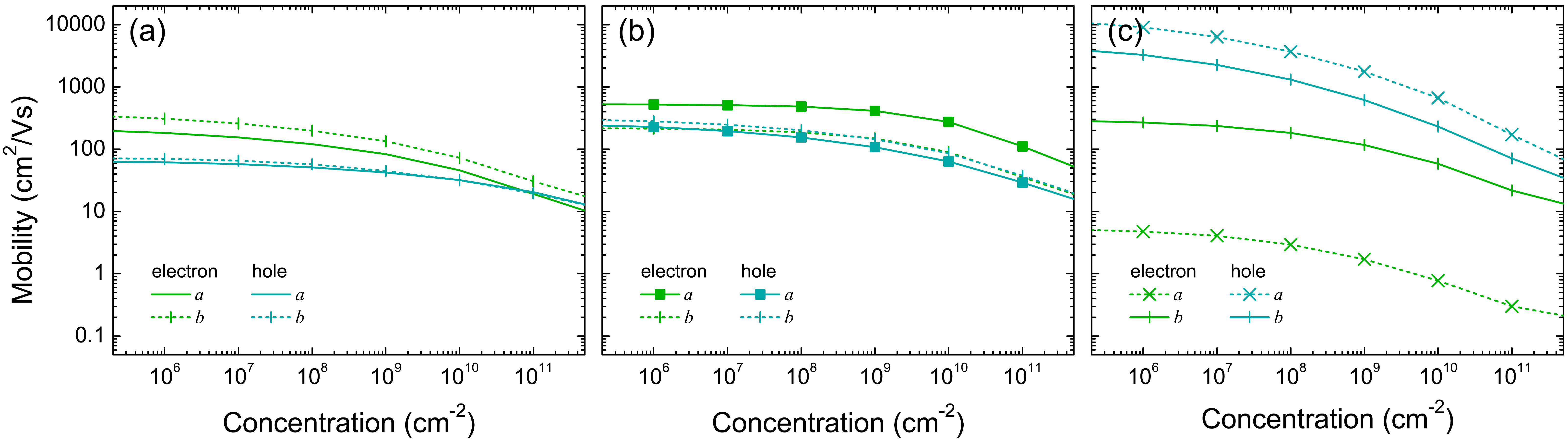}
\caption{
Mobility of monolayer (a) qTP1, (b) qTP2, and (c) qHP C$_{60}$ at 300 K as a function of carrier concentration.
}
\label{mobility} 
\end{figure*}

To confirm the transport properties, the carrier mobilities of all three phases at 300 K are calculated as a function of carrier concentration. As shown in Fig.\,\ref{mobility}, the mobilities for both electrons and holes decrease with increasing carrier concentration in all three phases, as ionized impurity scattering becomes stronger. Although the experimental carrier concentration is unknown, the calculated electron mobility along $a$ at low carrier concentrations ($< 10^9$ cm$^{-2}$), $1.7-4.8$ cm$^2$/Vs, is in perfect agreement with the measured electron mobility\,\cite{Hou2022}.

The obtained electron mobility for qTP1 C$_{60}$ in Fig.\,\ref{mobility}(a) is higher than the hole mobility in a wide doping range, in consistent with the more dispersive CB1 in Fig.\,\ref{electron}(a). For both electrons and holes, the mobility along the 1D chain ($a$ direction) is higher than that perpendicular to the chain ($b$ direction). For qTP2 C$_{60}$ in Fig.\,\ref{mobility}(b), the electron mobility along $a$ is the highest. This is unsuprising because the CB1 along $\Gamma$-X and CB2 along S-Y are more dispersive than other bands in Fig.\,\ref{electron}(b) and the CB states along $a$ in Fig.\,\ref{electron}(e) tend to overlap across the vertical [2+2] cycloaddition bonds. For qHP C$_{60}$, the hole mobilities along both directions are much higher than the electron mobilities, as shown in Fig.\,\ref{mobility}(c), which is in line with the dispersive VBM at $\Gamma$ along both directions in Fig.\,\ref{electron}(c) and the corresponding diffuse charge density in Fig.\,\ref{electron}(f). The electron mobility along $a$ is much lower than that along $b$ because the CBM along $\Gamma$-X is much flatter than that along Y-$\Gamma$. Despite that, even the lower bound of the mobility is still relatively high, as the non-localized $\pi$ bonds in C$_{60}$ allows efficient transfer of the photogenerated carriers\,\cite{Pan2020}.




\subsection*{Optical absorption}

\begin{figure*}[h]
\centering
\includegraphics[width=\textwidth]{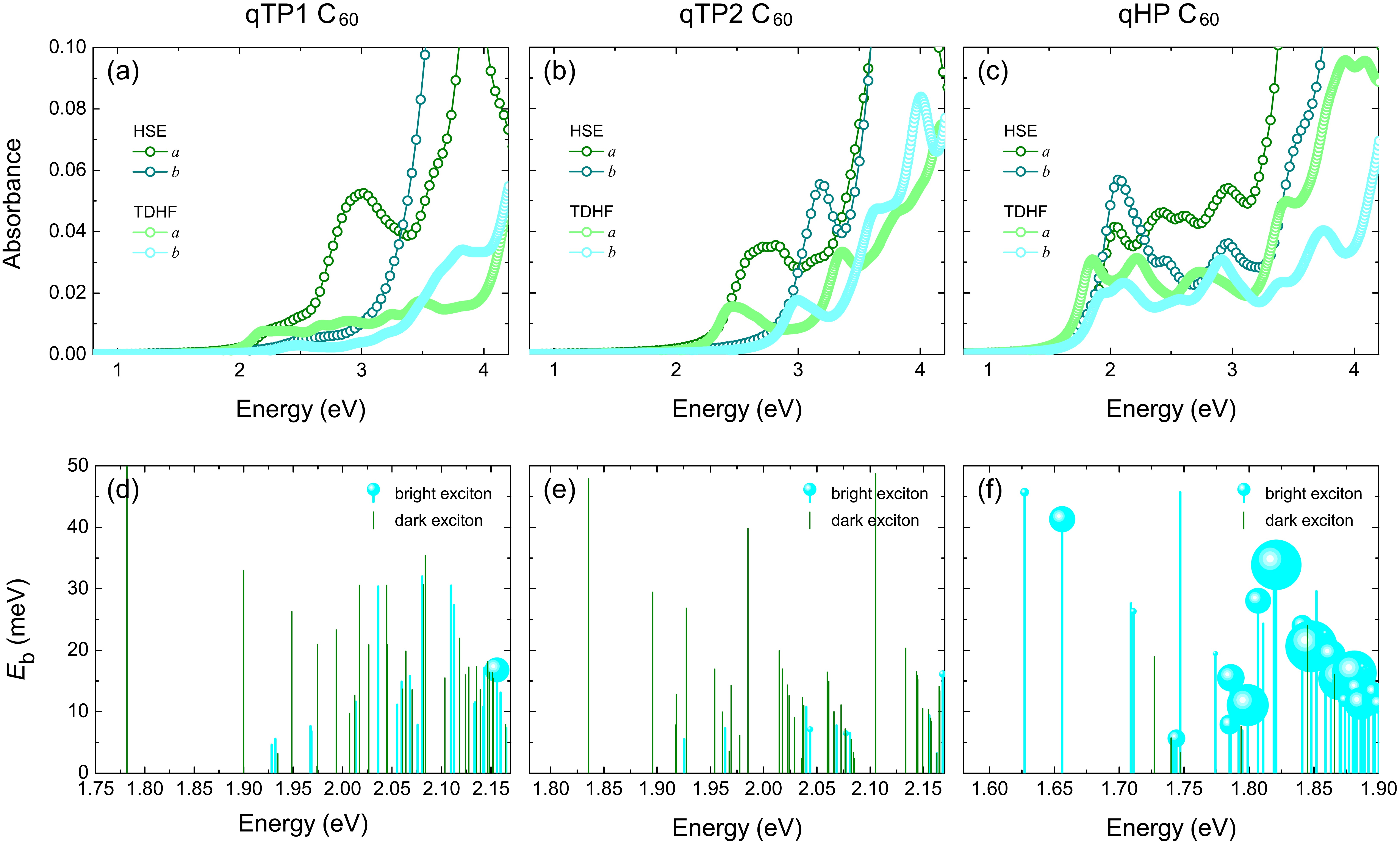}
\caption{
Imaginary part of the dielectric function $\epsilon_2$ of (a) qTP1, (b) qTP2, and (c) qHP C$_{60}$ calculated with HSE ($\mu = 0.11$ \AA$^{-1}$) and TDHF on top of HSE, as well as the binding energy $E_{\rm b}$ of the low-energy excitons in (d)-(f). The radius of the bright excitons indicates the oscillator strength. The larger the radius, the higher the oscillator strength.
}
\label{exciton} 
\end{figure*}

Having established that all three fullerene networks can separate the carriers effectively in 2D, their absorption spectra for photocatalysis are then investigated. The thickness-independent absorbance $A (\omega)$ of monolayer fullerene networks is first calculated by using the weakly screened hybrid functional with $\mu = 0.11$ \AA$^{-1}$, corresponding to the optical absorption of the hybrid-functional electronic structures in the independent particle picture. The absorbance of all three phases is gathered in Fig.\,\ref{exciton}(a)-(c). Within the independent particle approximation, the low-energy absorbance of both qTP1 and qTP2 C$_{60}$ is strongly anisotropic along the $a$ and $b$ directions, whereas the first absorbance peaks of qHP C$_{60}$ have similar energies along both directions. Moreover, the indirect band gaps of qTP1 and qTP2 C$_{60}$, along with the low optical transition probabilities between the highest valence and lowest conduction bands, give rise to low optical absorbance in the visible light region.

Beyond the independent particle approximation, the absorbance is evaluated by HSE + TDHF to assess the excitonic contributions, as demonstrated by the green and cyan curves in Fig.\,\ref{exciton}(a)-(c). In monolayer qTP1 C$_{60}$, including excitonic effects leads to much weaker optical absorbance, as shown in Fig.\,\ref{exciton}(a). This is because the low-energy excitons are mostly dark and the optical transitions involved in these dark excitons have zero oscillator strengths, as demonstrated in Fig.\,\ref{exciton}(d). For monolayer qTP2 C$_{60}$, although the oscillator strengths in the low-energy range are mostly zero in Fig.\,\ref{exciton}(e), the exciton absorbance peak in monolayer qTP2 C$_{60}$ is only moderately weaker than the independent particle absorbance peak in Fig.\,\ref{exciton}(b). Compared to its qTP counterparts, much stronger exciton absorbance peaks are observed in monolayer qHP C$_{60}$, as shown in Fig.\,\ref{exciton}(c). The low-energy excitons in monolayer qHP C$_{60}$ are mostly bright with binding energies around $5-50$ meV, as present in Fig.\,\ref{exciton}(f). Therefore, strong exciton absorbance is induced in qHP C$_{60}$, and in particular, the absorbance around 2 eV ($0.20-0.32$) is even stronger than those in zero band-gap graphene\,\cite{Bechstedt2012,Matthes2016} and in large band-gap photocatalysts such as monolayer GaSe\,\cite{Zhuang2013a} and blue phosphorus/Mg(OH)$_2$ van der Waals heterostructures\,\cite{Wang2018a}, which makes qHP C$_{60}$ a promising photocatalytic material to effectively utilize the solar spectrum around 2 eV.

\subsection*{Band alignment}


 
\begin{figure*}
\centering
\includegraphics[width=\textwidth]{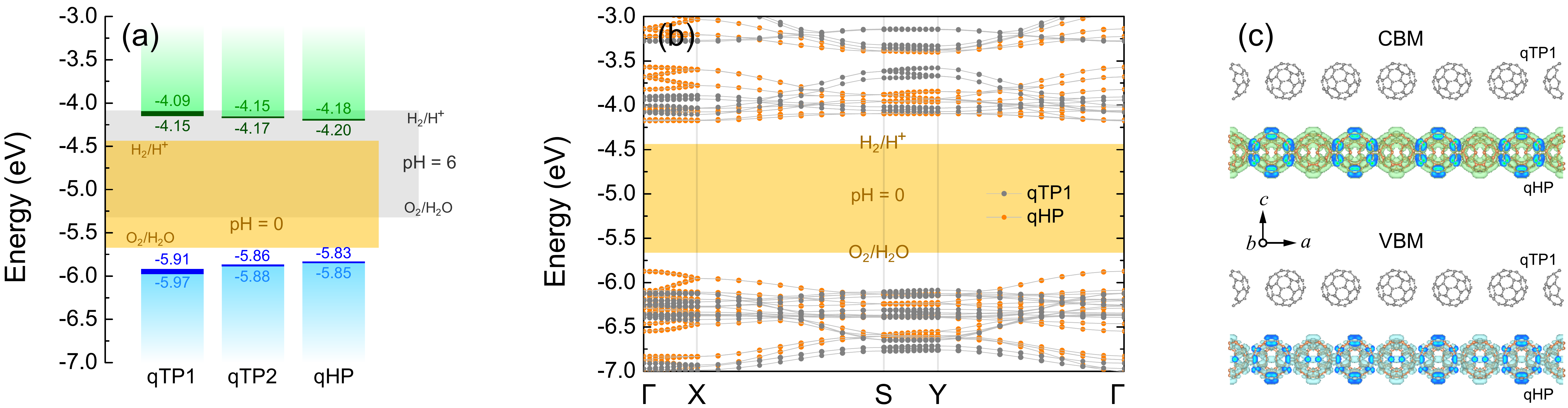}
\caption{
(a) Band alignment of qTP1, qTP2 and qHP C$_{60}$ monolayers calculated with HSE ($\mu = 0.11$ \AA$^{-1}$). The CBM and VBM positions in the independent particle picture are marked in green and cyan respectively, while the CBM and VBM positions including the excitonic effects are marked in dark green and blue respectively. (b) Band structures of qTP1/qHP heterostructures, with the vacuum level set to zero. (c) Partial charge density of the CBM and VBM states in the qTP1/qHP heterostructures. The default isosurface level (0.002 \AA$^{-3}$) is used, as implemented in {\sc vesta}\,\cite{vesta}.
}
\label{alignment} 
\end{figure*}

The exciton absorbance peaks in monolayer C$_{60}$ networks around 2 eV can maximize the solar energy absorption for water splitting\,\cite{LeBahers2014,Brlec2022}. For an overall water splitting reaction, the energy levels of the CBM and VBM must straddle the redox potentials of water. In other words, the CBM (with respect to the vacuum level) should be higher than the hydrogen evolution potential of $-4.44 + {\rm pH} \times 0.059$ eV, while the VBM should be lower than the oxygen evolution potential of $-5.67 + {\rm pH} \times 0.059$ eV\,\cite{Chakrapani2007,Zhang2019d,Chen2022}. To determine the band edge positions of qTP1, qTP2 and qHP C$_{60}$ monolayers, the vacuum levels of all three phases are calculated by averaging the electrostatic potential along the $c$ axis. Figure\,\ref{alignment}(a) summarizes the HSE band alignment of all three C$_{60}$ monolayers with $\mu = 0.11$ \AA$^{-1}$ (for band alignment calculated with PBEsol, HSEsol and unscreened hybrid functional, see the Supporting Information). 
In monolayer qTP1 C$_{60}$, the CBM is 0.35 eV higher than the reduction reaction potential of H$_2$/H$^+$ at pH = 0, which is suitable for water reduction. Moreover, the VBM is 0.30 eV lower than the oxidation potential of O$_2/$H$_2$O at pH = 0, which is suitable for water oxidation. Similarly, the CBM of qTP2 C$_{60}$ is 0.29 eV higher than the reduction potential and the VBM is 0.22 eV lower than the oxidation potential. Regarding monolayer qHP C$_{60}$, the CBM lies 0.26 eV above the reduction potential and the VBM is 0.18 eV below the oxidation potential. Including the exciton binding energy leads to band edge shifts towards the redox potential by 0.06 eV for qTP1 C$_{60}$, while the band edge shifts in qTP2 and qHP C$_{60}$ are about 0.02 eV. Therefore, all three C$_{60}$ monolayers exhibit large band gaps with appropriate band edge positions for overall photocatalytic water splitting at pH = 0. Increasing the pH upshifts the redox potentials of water, and at pH = 6, all three phases of monolayer C$_{60}$ are no longer suitable for water reduction.

The lattice parameters of $3\times1$ qTP1 C$_{60}$ and $2\times1$ qHP C$_{60}$ are matched within 3.8\% for $a$ and 0.8\% for $b$ respectively, therefore monolayer qTP1 and qHP C$_{60}$ can also be combined to form qTP1/qHP heterostructures. To identify the type of the heterostructures for device applications, the band alignment at the qTP1/qHP interface is investigated. Compared to qTP1 C$_{60}$, qHP C$_{60}$ has a consistently smaller band gap, as shown in Fig.\,\ref{alignment}(a). The offset between the conduction band edges of qTP1 and qHP C$_{60}$ monolayers is 0.09 eV with the CBM of qHP lower than that of qTP1, and a higher VBM of qHP relative to qTP1 leads to a valence band discontinuity of 0.12 eV. Consequently, a type-I (straddling gap) band alignment exists between qTP1 and qHP C$_{60}$. Geometry optimization of the qTP1/qHP heterostructures results in 3.5\% strain along $a$ and 0.3\% strain along $b$ for qTP1 C$_{60}$, while compresses the qHP C$_{60}$ lattice by 0.4\% and 0.5\% along $a$ and $b$ respectively (for strain effects on band alignment of individual monolayers, see the Supporting Information). Despite that, the band alignment is still type-I, as demonstrated in Fig.\,\ref{alignment}(b). The type-I heterostructures with qTP1 and qHP C$_{60}$ can be utilized in optical devices such as light-emitting diodes owing to high emission efficiency\,\cite{Zheng2018}, or in lasers because of efficient recombination of spatially confined electrons and holes\,\cite{Ozcelik2016}. As confirmed by the partial charge density of CBM and VBM in Fig.\,\ref{alignment}(c), these states are confined in monolayer qHP C$_{60}$.

\subsection*{Thermodynamic driving force for water splitting}

The thermodynamics of water adsorption on monolayer fullerene networks are investigated by calculating the total energy difference between the H$_2$O-adsorbed C$_{60}$ and individual systems (i.e. pristine monolayer C$_{60}$ and isolated H$_2$O molecule)\,\cite{Ju2020}. The obtained adsorption energies for qTP1, qTP2 and qHP C$_{60}$ are -0.151, -0.109 and -0.107 eV respectively, indicating their capability of water adsorption.

\begin{figure*}[h]
\centering
\includegraphics[width=\textwidth]{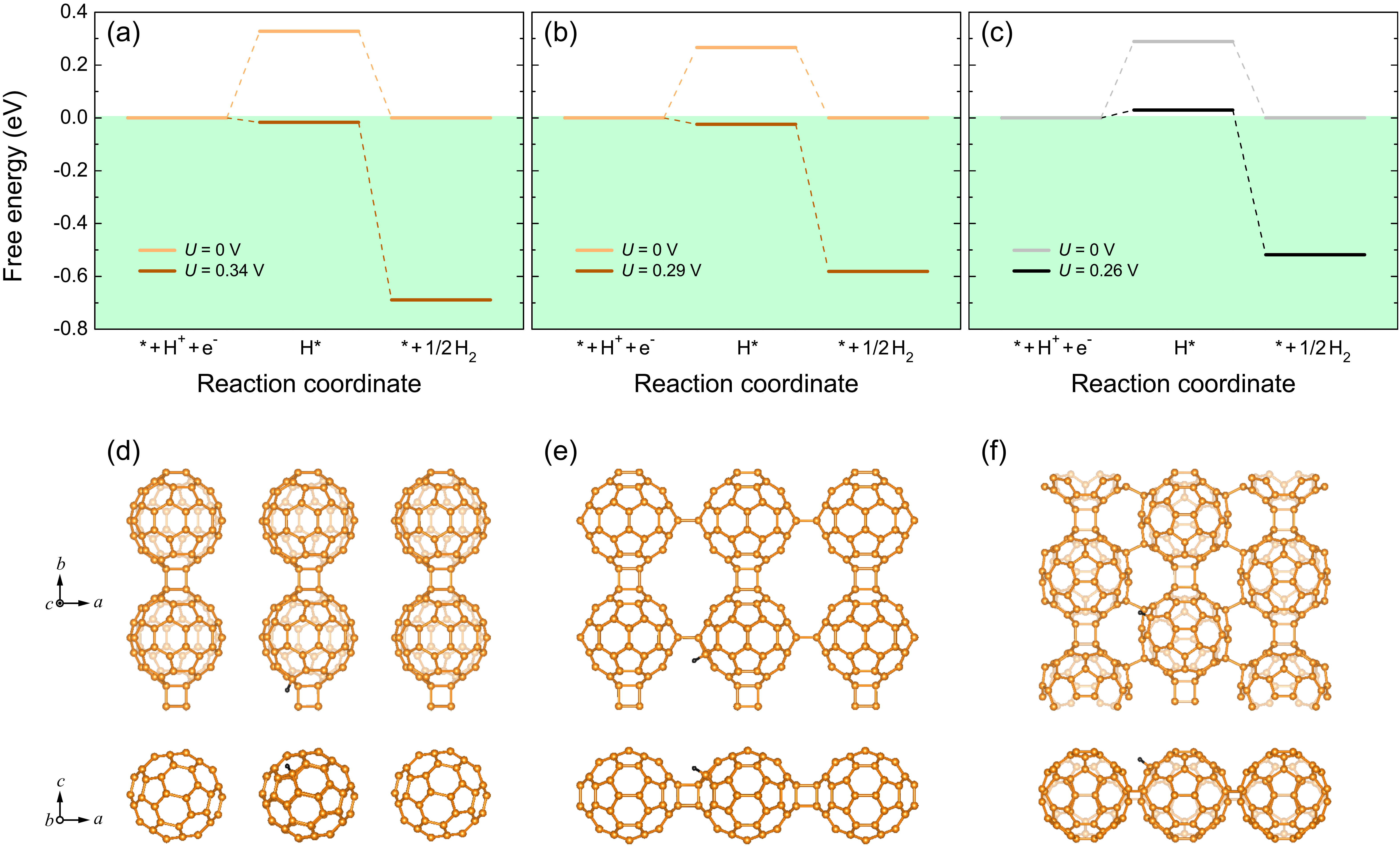}
\caption{
Free-energy diagram for hydrogen reduction reaction at pH = 0 and room temperature in (a) qTP1, (b) qTP2, and (c) qHP C$_{60}$, with the Gibbs free energy of the combination of monolayer fullerene networks, a proton and an electron set to zero. $U=0$ V corresponds to the absence of photoexcitaion. The nonzero potential $U$ is generated by photoexcited electrons in the CBM. The lowest energy intermediates H* for all three phases are present in (d)-(f).
}
\label{her} 
\end{figure*}

The thermodynamics of the hydrogen evolution reaction are investigated by calculating the Gibbs free energy of the intermediates of the reaction at pH = 0 and room temperature\,\cite{Norskov2004,Rossmeisl2007} (for details on the half-reaction of water oxidation, see the Supporting Information). As shown in Fig.\,\ref{her}(a)-(c), the hydrogen evolution reaction has two steps. In the first step, monolayer fullerene networks (denoted as *) combine with a proton (H$^+$) and an electron (e$^-$) to form H* species. In the second step, H$_2$ molecules are formed from the H* species. The lowest energy intermediates H* for all three phases are present in Fig.\,\ref{her}(d)-(f). For qTP1 C$_{60}$, the hydrogen atom is adsorbed at the top site of the nearest neighboring carbon atom to the [2+2] cycloaddition bonds. Similarly, the adsorbed H atom on qTP2 C$_{60}$ is at the top site of the nearest neighboring carbon atom to the vertical [2+2] cycloaddition bonds. Different from qTP1 and qTP2 C$_{60}$, in the H-adsorbed qHP C$_{60}$ a C$-$H bond is formed between the hydrogen atom and the second nearest carbon atom to the C$-$C single bond.

In the absence of photoexcitation ($U=0$ V), all three phases of monolayer fullerene networks, when forming the lowest energy H* species, exhibit unfavorable positive Gibbs free energies (0.327, 0.266 and 0.289 eV for qTP1, qTP2 and qHP C$_{60}$ respectively). Then the release of H$_2$ molecules from the H* species is exothermic. Upon light irradiation, the photoexcited electrons in the CBM generate an external potential $U$ of 0.345, 0.291 and 0.259 eV for qTP1, qTP2 and qHP C$_{60}$ respectively, corresponding to the potential difference between the CBM and the H$_2$/H$^+$ reduction potential. Consequently, both steps in the hydrogen reduction reaction (the formation of H* species and the release of H$_2$ molecules) in the free-energy diagram are downhill for qTP1 and qTP2 C$_{60}$. Therefore, both qTP1 and qTP2 C$_{60}$ can efficiently split water under acidic environment upon light irradiation as the hydrogen reduction reaction can spontaneously proceed. Regarding qHP C$_{60}$, the reaction barrier is significantly reduced to 0.030 eV under photoexcitation, which is close to the thermal fluctuation energy $k_BT$ at room temperature (0.026 eV). In addition, it has been reported that the experimentally obtained qHP C$_{60}$ flakes tend to be negatively charged\,\cite{Hou2022,Gottfried2022}, which can provide further external potential for hydrogen evolution reaction.

\subsection*{Discussion}

Monolayer fullerene networks can be combined with a highly diverse set of lattice-matched 2D materials with higher CBM and VBM\,\cite{Ozcelik2016,Zhang2016e,Hu2017,Yang2020a} to form type-II heterostructures to separate electrons and holes in individual layers, which can further improve the photocatalytic performance (for type-II band alignment of qTP2/SnTe and qTP2/PbTe heterostructures, see the Supporting Information). The presence of monolayer fullerene networks can improve the separation of electrons and holes by trapping them individually into different nanostructures, i.e. 0D C$_{60}$ cages in all three phases, or 1D C$_{60}$ chains in qTP1 fullerene. For the 0D C$_{60}$ cages in all three phases, the non-localized $\pi$ bonds in C$_{60}$ allows continuous transfer and separation of the photogenerated carriers\,\cite{Pan2020}. 
Furthermore, the enhanced surface area in monolayer fullerene networks, with more micropores and surface active sites compared to any 2D materials, can significantly increase the photocatalytic efficiency. Additionally, the optical transition oscillator strength in both the qTP1 and qTP2 monolayers is quite low, thereby suppresses the carrier recombination and enhances the photocatalytic efficiency as an electron acceptor. Regarding monolayer qHP C$_{60}$, the strong optical absorbance can generate a large amount of electrons, making it promising for providing electrons for hydrogen evolution.

Most interestingly, fullerene itself, after doping\,\cite{Zhao2005,Durbin2016} or coating\,\cite{Yoon2008,Wang2009}, can act as promising molecular hydrogen attractors. Theoretical calculations have reported that, one transition metal atom bound to fullerene can bind eleven hydrogen atoms, with a binding energy of 0.3 eV that is ideal for vehicular applications because of its ability to adsorb and desorb H$_2$ reversibly\,\cite{Zhao2005}. In addition, the maximum hydrogen storage density can reach $6-9$ wt \% near ambient pressure at room temperature\,\cite{Zhao2005,Yoon2008,Wang2009}, which is highly desirable for fuel-cell powered vehicles. Moreover, there is both theoretical and experimental evidence that fullerene can be decorated with various metal atoms while remaining stable\,\cite{Tast1996,SankarDe2018}. In monolayer fullerene networks, the decorating atoms can be uniformly distributed to form monolayer coating, which may further increase the retrievable hydrogen storage density.

\section*{Conclusion}

In summary, I use weakly screened hybrid functional to examine the band structures of monolayer C$_{60}$, rationalizing the measured electronic band gap. On top of the electronic structures, time-dependent Hartree-Fock calculations predict excellent exciton binding energy, reproducing the measured optical band gap. To gain insights into the photocatalytic performance of monolayer fullerene networks, I investigate the band alignment of monolayer fullerene networks, and find that all three phases have the band edge positions suitable for overall water splitting. The overall water splitting can occur spontaneously under acidic condition at room temperature upon photoexcitation. The distinct optical properties between qTP and qHP fullerene provide unique advantages for different applications in photocatalysis, with qTP C$_{60}$ being a likely electron acceptor and qHP C$_{60}$ being a promising electron donor respectively. Beyond water splitting, the type-I band alignment for the qTP1/qHP heterostructures offers new opportunities for optical devices and lasers.


\section*{Acknowledgement}

I thank Prof. Bartomeu Monserrat and Dr. Ivona Bravi\'{c} at the University of Cambridge for helpful discussions. I acknowledge support from the Winton Programme for the Physics of Sustainability, and from Magdalene College Cambridge for a Nevile Research Fellowship. The calculations were performed using resources provided by the Cambridge Tier-2 system, operated by the University of Cambridge Research Computing Service (www.hpc.cam.ac.uk) and funded by EPSRC Tier-2 capital grant EP/P020259/1, as well as with computational support from the U.K. Materials and Molecular Modelling Hub, which is partially funded by EPSRC (EP/P020194), for which access is obtained via the UKCP consortium and funded by EPSRC grant ref. EP/P022561/1.

\bibliography{references}

\end{document}